\documentclass[lettersize,journal]{IEEEtran}

\usepackage{svg}
\usepackage[numbers]{natbib}
\usepackage{float}
\usepackage{hyperref}
\usepackage{pdfpages}
\usepackage{soul}
\usepackage{stfloats}
\usepackage{balance}

\definecolor{cyan}{RGB}{68,114,196}

\ifCLASSINFOpdf\else\fi

\usepackage{adjustbox,lipsum}
\usepackage{titlesec}          
\usepackage{algorithm}
\usepackage{algpseudocode}
\usepackage{amsmath,amssymb}
\usepackage{booktabs}
\usepackage{multirow}
\usepackage{multicol}
\usepackage{graphicx}
\usepackage{listings}
\usepackage{xurl}                 
\hypersetup{breaklinks=true}      
\begin{document}

\title{API Security Based on Automatic OpenAPI Mapping}

\author{Yarin~Levi%
\thanks{Yarin Levi is with the Department of Computer Science and the Ariel Cyber Innovation Center, Ariel University, Ariel, Israel (e-mail: yarin.levi4@msmail.ariel.ac.il).}%

\and Ran~Dubin%
\thanks{Ran Dubin is with the Department of Computer and Software Engineering and the Ariel Cyber Innovation Center, Ariel University, Ariel, Israel (e-mail: rand@ariel.ac.il).}%
}

\markboth{IEEE TRANSACTIONS ON INFORMATION FORENSICS AND SECURITY}%
{Levi and Dubin: API Security Based on Automatic OpenAPI Mapping}

\maketitle

\begin{abstract}
This paper presents Map Reduce Graph (MRG), a novel unsupervised method for modeling and securing HTTP REST APIs. MRG learns API structure from real-world traffic without prior knowledge or labels, automatically generating OpenAPI-compliant documentation by reconstructing routes, methods, and parameter formats.

MRG enables real-time updates, explainable visualization, and anomaly detection, helping identify undocumented or evolving behaviors. It detects malformed requests, structural deviations, and injection attacks using graph-based validation and a deep autoencoder for payload analysis. Compared to state-of-the-art methods like HRAL and FT-ANN, MRG achieves up to 11.4\% higher recall, over 20 times faster inference, and perfect precision (100\%) on multiple API-layer attacks.

Designed for dynamic microservice environments, MRG operates in three phases—training, updating, and detection—and integrates smoothly with observability and security tools. This work contributes a fully automated, efficient pipeline for real-time API visibility, schema inference, and anomaly detection without manual tuning or labeled data.
\end{abstract}


\section{Introduction}
In today's digital landscape, Application Programming Interfaces (APIs) have become the backbone of modern software ecosystems. They facilitate seamless interactions between microservices, mobile applications, cloud services, and IoT devices, enabling everything from payment processing and user authentication to telemetry collection and machine learning workflows. As of 2023, API calls constituted 71\% of all internet traffic, highlighting their central role in digital communications \cite{imperva2024api}.

However, this exponential growth in API usage has introduced significant security and observability challenges. APIs are often exposed externally, evolve rapidly, and suffer from incomplete or outdated documentation. According to a recent study, 84\% of security professionals experienced at least one API-related incident in the past year, with average remediation costs of \$591,404 in the United States, and costs rising to \$832,801 in the financial services sector \cite{akamai2023api}. Globally, organizations are losing between \$94 billion and \$186 billion annually due to vulnerable or insecure APIs and automated abuse by bots \cite{imperva2024economic}.

Real-world incidents underscore these risks. In January 2023, T-Mobile disclosed that the personally identifiable information of 37 million customers had been breached through an API attack \cite{apnews2023tmobile}. Similarly, the Optus data breach in September 2022 exposed sensitive information of up to 10 million customers due to an unprotected and publicly exposed API \cite{upguard2022optus}. In May 2021, Peloton experienced a data breach where an API vulnerability allowed unauthorized access to user data, highlighting the lack of proper authentication and authorization measures \cite{twingate2021peloton}.

Despite growing threats, traditional security mechanisms such as Web Application Firewalls (WAFs) and Intrusion Detection Systems (IDS) are often inadequate for securing modern APIs. These systems rely heavily on static signatures or predefined rules, rendering them ineffective against novel, obfuscated, or logic-based attacks, particularly in dynamic API environments where endpoints evolve rapidly and documentation is incomplete or missing.

Machine learning-based approaches also fall short, as they typically treat API requests as flat token sequences, failing to model the hierarchical structure of RESTful APIs, such as nested paths, dynamic parameters, and method-path relationships. This limits their ability to detect structural anomalies, undocumented endpoint usage, or subtle deviations from normal API behavior.

To overcome these limitations, API-specific security solutions must understand and enforce the API traffic's structural and semantic context. These solutions can be deployed either \textbf{inline}, for real-time enforcement and request blocking, or \textbf{out-of-band}, for passive monitoring, retrospective analysis, and detection of complex or delayed threats.

To address these limitations, we propose \textbf{Map Reduce Graph (MRG)}—a novel, unsupervised method for learning and modeling the structure of HTTP REST APIs directly from live traffic. MRG automatically generates accurate OpenAPI specifications without requiring labeled data or prior knowledge of the existing API endpoint documentation. The framework operates in three distinct phases:
\begin{itemize}
    \item \textbf{Training:} Analyzes incoming HTTP requests to extract and aggregate structural patterns.
    \item \textbf{Updating:} Incorporates new request patterns over time, supporting continuous learning.
    \item \textbf{Anomaly Detection:} Identifies deviations from the learned structure, flagging potentially malicious or anomalous behavior.
\end{itemize}

MRG introduces a tree-based, map-reduce approach where API endpoints are represented as hierarchical trees. The mapping phase decomposes paths and parameters, while the reduction phase generalizes them into a canonical form using placeholders (e.g., \texttt{\{id\}}, \texttt{\{param\}}). This structure is further modeled as a graph, capturing both local and global relationships among request components. Unlike vector-based models, this graph representation allows for precisely detecting structural anomalies and unusual patterns, such as nested path manipulation or unexpected query combinations.

By automatically learning all API usage, dynamic specification inference, and anomaly detection in a single system, MRG bridges the gap between API observability and security. It offers competitive performance in structured settings and demonstrates fast inference times, though further improvements are needed to generalize across less-structured environments. This paper details the design, implementation, and evaluation of MRG and demonstrates its advantages through experiments on real-world and synthetic API datasets.

The paper is structured as follows: \textbf{Section~\ref{sec:contribution}} outlines the key contributions of our work, including the introduction of the MRG framework. \textbf{Section~\ref{sec:relatedwork}} reviews relevant prior research in API security, automatic documentation, and anomaly detection, positioning our approach within the current body of knowledge. \textbf{Section~\ref{sec:methodology}} describes the architecture and detailed workflow of the MRG framework, including tree construction, graph transformation, and anomaly detection mechanisms. \textbf{Section~\ref{sec:graphdetection}} elaborates on the graph-based anomaly detection strategy, discussing schema generalization, graph comparison, and advantages over traditional methods. \textbf{Section~\ref{sec:datasets}} introduces the datasets used for evaluation, including CSIC 2010, ATRDF, and the proprietary ATRDF2 dataset. \textbf{Section~\ref{sec:evaluation}} details our evaluation methodology, comparing MRG against baseline approaches using standard metrics. Moreover, it presents experimental results, providing quantitative comparisons and analysis across multiple datasets. \textbf{Section~\ref{sec:limitations-future-work}} discusses the limitations of the current framework and suggests directions for future improvements. Finally, \textbf{Section~\ref{sec:conclusion}} summarizes our findings and concludes the paper.

\section{Contribution}
\label{sec:contribution}
\begin{figure}[!t]
    \centering
    \includegraphics[width=\linewidth]{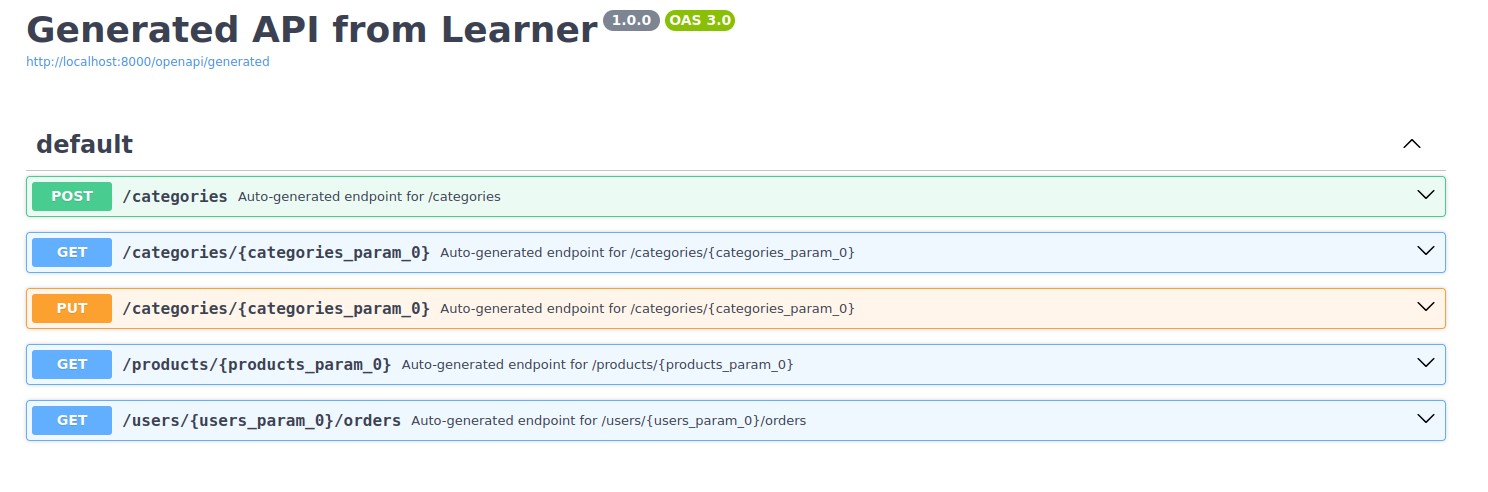}
    \caption{Automatically generated OpenAPI UI from MRG showing endpoints and parameterized paths inferred from traffic.}
    \label{fig:openapi_ui}
\end{figure}

\begin{figure*}[!t]
    \centering
    \includegraphics[width=\textwidth]{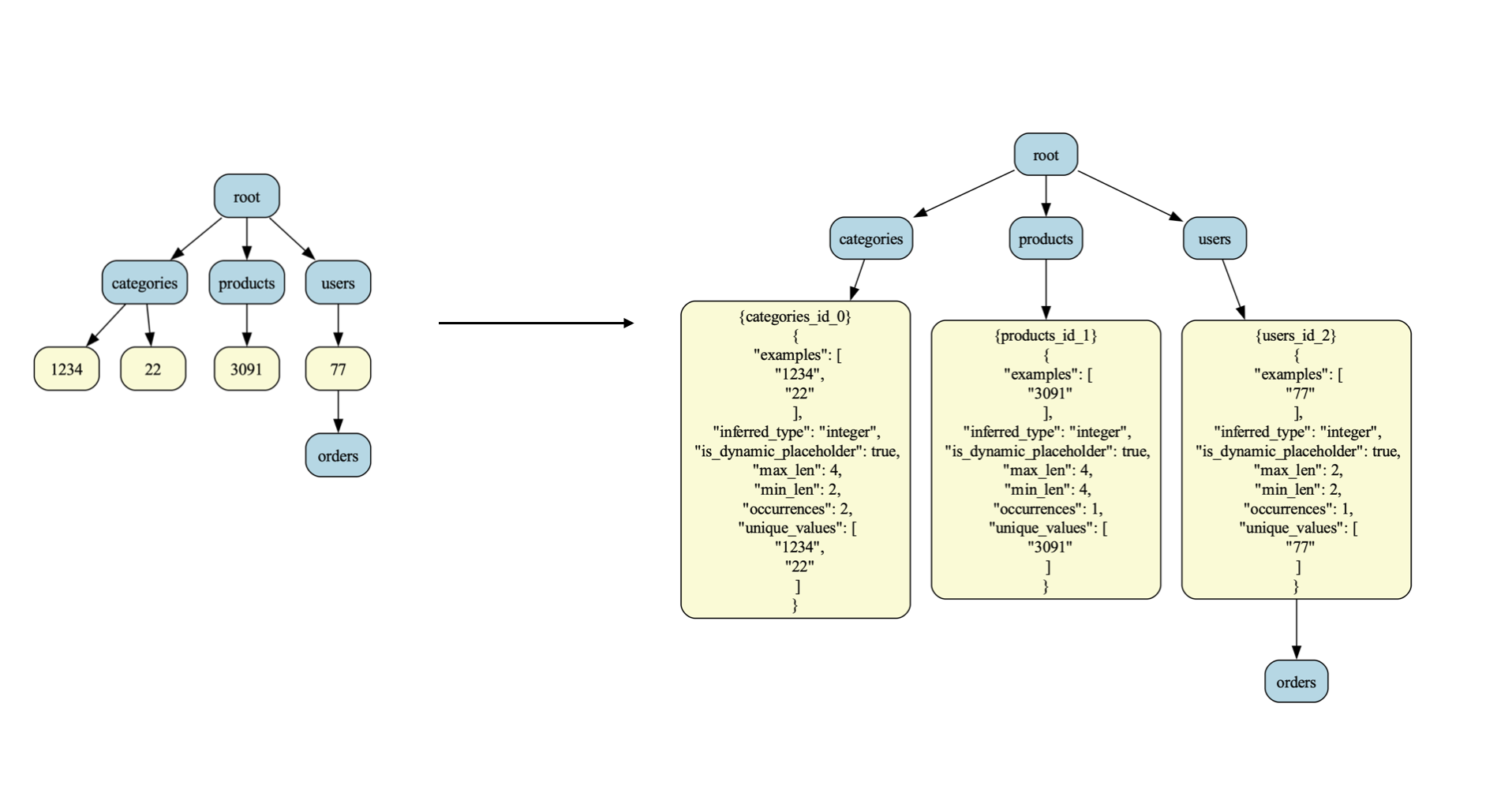}
    \caption{Tree structure before and after reduction. This visual illustrates how API paths are generalized using placeholders.
    The left side presents a raw API tree structure as learned from observed traffic. Each node corresponds to a URL path segment, including dynamic values such as user IDs or order IDs, which are represented explicitly. The right side shows the reduced tree schema after applying the generalization phase. Dynamic segments (yellow nodes in the tree) are merged into placeholders (e.g., \texttt{\{categories\_id\}}) based on inferred types and structural patterns. This process abstracts variable components while preserving essential relationships, significantly reducing complexity. This simplified schema forms the foundation for API specification generation and anomaly detection in MRG.}
    \label{fig:tree_structure_fullwidth}
\end{figure*}

We offer several notable contributions to the field of API security and observability:
\begin{itemize}
    \item \textbf{MRG: A Novel Unsupervised Tree-Graph Framework for API Schema Learning:} We introduce MRG, a unique unsupervised method for learning and modeling HTTP REST API structures directly from live traffic (see visualization in ~\ref{fig:tree_structure_fullwidth}). This framework operates in three phases: the Map phase decomposes HTTP requests into hierarchical tree structures and graph representations; the Reduce phase aggregates patterns and generalizes dynamic segments into canonical placeholders (e.g., \texttt{\{id\}}), forming a reusable schema; and the Detect phase leverages this learned graph-based API schema to identify structural anomalies and potential malicious traffic.

    \item \textbf{Automated and Explainable OpenAPI Specification Generation:} Our system significantly enhances API visibility by automatically generating detailed and human-readable OpenAPI specifications via OpenAPI UI from observed traffic. This includes continuous updates to reflect evolving API behaviors and the automatic identification of both documented and previously undocumented (shadow) APIs. The generated specifications are designed to be highly explainable, providing security analysts and developers with clear, up-to-date documentation for reliable validation and enforcement. {Figure~\ref{fig:openapi_ui} displays the rendered OpenAPI UI showing the set of endpoints, methods, and generalized parameter paths (e.g., \texttt{/user/\{id\}/orders}) inferred by MRG from real-world traffic, enabling transparent schema inspection.}

    \item \textbf{Visibility \& Reset Options:} The system supports streaming input, adapting incrementally to new API patterns without requiring full retraining. It provides real-time visualization of the inferred API tree structure, offering security analysts immediate insights into endpoint behaviors. Additionally, the system includes a reset mechanism, enabling analysts to easily re-baseline the model for environmental changes or testing.

    \item \textbf{New Dataset - API Traffic Research Dataset Framework v2 (ATRDF2):} We introduce ATRDF2 \cite{ATRDF2}, a novel synthetic dataset containing a large volume of both normal and anomalous API traffic. This dataset is specifically designed to evaluate generalization and robustness against diverse, LLM-generated API-layer attacks, facilitating future research.
\end{itemize}

\section{Related Work}
\label{sec:relatedwork}

Traditional WAFs and IDSs, like Snort~\cite{roesch1999snort}, rely on predefined signatures, which quickly become outdated and struggle with dynamic API environments, as shown by their inability to detect up to 89\% of JSON-based SQL injection payloads in recent tests~\cite{balasys2022waf}. This limitation stems from their reliance on exact byte pattern matching, which fails to adapt to constantly changing API paths, parameters, and encodings.

While machine learning approaches, such as CNN/LSTM hybrids~\cite{zhao2024multi}, Markov models~\cite{kruegel2003web}, and GNNs~\cite{du2022tsgnn}, have been applied to HTTP traffic anomaly detection, they face two key challenges~\cite{diazverdejo2023review}:
\begin{enumerate}
    \item Most models flatten hierarchical URLs into token sequences, losing crucial structural information and context~\cite{APICDR}.
    \item They typically require large, labeled datasets for training, which are scarce in security applications.
\end{enumerate}
FT-ANN~\cite{Aharon2024} introduces a novel few-shot anomaly detection framework based on FastText embeddings and Approximate Nearest Neighbor (ANN) search. It incorporates a custom tokenizer tailored for API syntax and emphasizes efficient detection through a classification-by-retrieval approach. Despite its strong performance on benchmark datasets like CSIC and ATRDF, FT-ANN still relies on vector encodings and a predefined training set. In contrast, MRG leverages a graph-based representation that preserves the full structure of API requests and learns directly from traffic in an unsupervised manner—eliminating the need for labeled data while capturing richer semantic and structural information.

Finally, detecting API logic flaws (e.g., IDOR, Broken Authentication, Excessive Data Exposure), which are high on the OWASP API Security Top 10~\cite{owasp_api_top10}, is challenging as they involve incorrect behavior, not invalid input. Tools like APISpec~\cite{hu2022apispec} detect violations by comparing runtime behavior to OpenAPI specs, and RESTler~\cite{atlidakis2019restler} uses fuzzing. However, these require accurate and complete documentation, often unavailable. MRG differs by learning the API structure directly from live traffic and encoding it into a graph schema. This schema captures method-to-path bindings, parameter types, dynamic IDs, and header requirements, enabling detection of unauthorized access, missing authentication, and method misuse—issues missed by traditional signature or payload-based tools. MRG's unified framework addresses both visibility and security without prior knowledge or labeled examples, distinguishing it from prior work.

Our work is compared against HTTP REST API Learning (HRAL)~\cite{APICDR}, FT-ANN~\cite{Aharon2024}, and traditional anomaly detection techniques (One-Class SVM, Isolation Forest, centroid-based classifiers), covering structural, embedding-based, and statistical approaches. Speculator was excluded due to its inferiority compared to HRAL.

\section{Methodology}
\label{sec:methodology}
In this section, we present the architecture and processing pipeline of the proposed system for learning and enforcing API schemas. The methodology is divided into two main phases: a learning phase that constructs a generalized schema from observed API requests, and an enforcement phase that validates incoming requests against this schema. We further detail each phase through a step-by-step description of the system's internal processes

\subsection{System Overview}
\begin{figure}[!t]
    \centering
    \includegraphics[width=\linewidth]{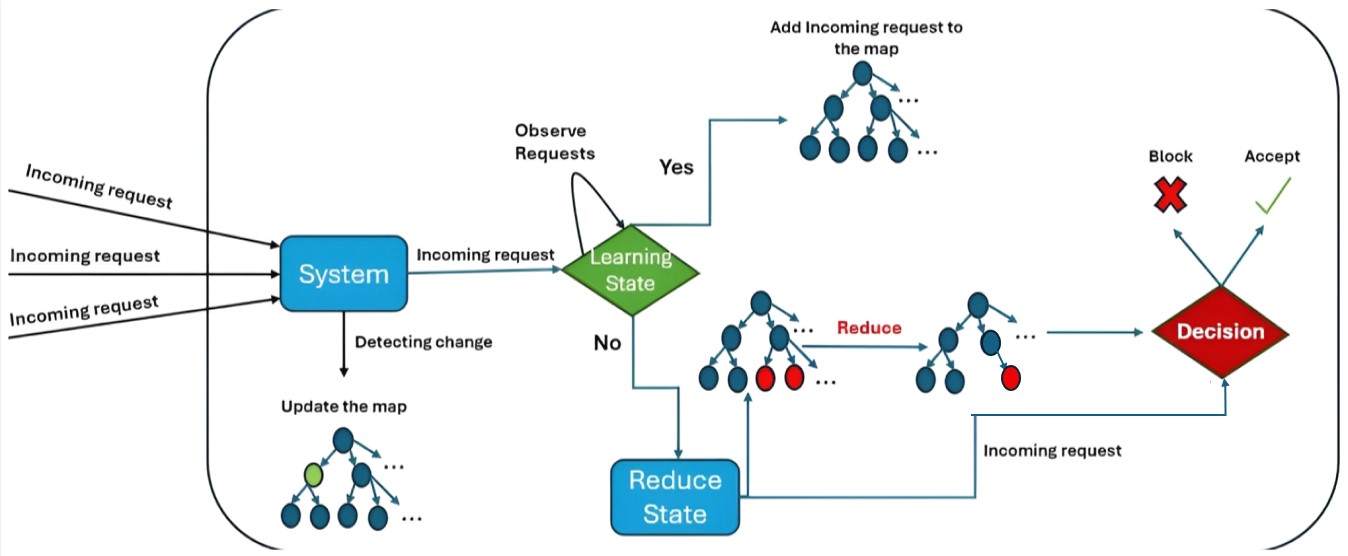}
    \caption{MRG Methodology Overview. This diagram outlines the MRG system's concise process: (1) \textbf{Ingest Traffic:} The system receives real-world HTTP traffic. (2) \textbf{Map Phase:} Requests are parsed into hierarchical trees in a "Learning State." (3) \textbf{Reduce Phase:} Trees are aggregated and generalized into a graph schema. (4) \textbf{Anomaly Detection:} Incoming requests are validated against the schema (structural) and payload content (semantic via autoencoder). (5) \textbf{Output/Action:} Detected anomalies trigger security alerts; requests are blocked/accepted; OpenAPI specs are inferred.}
    \label{fig:methodology_overview}
\end{figure}

Figure~\ref{fig:methodology_overview} illustrates the full MRG pipeline, from traffic ingestion and schema learning to anomaly detection and response:

\textbf{Learning (Map) Phase:}  
Incoming HTTP requests are parsed and decomposed into a hierarchical tree structure. Each node in this tree corresponds to a segment of the API path, while associated metadata (such as HTTP methods and query parameters) is captured. The tree is then transformed into a graph, where nodes represent URL segments and query parameters, and edges represent their connections. This graph encapsulates the overall structure of the API.

\textbf{Enforcing (Reduce) Phase:}  
After accumulating sufficient data based on the user's decision, the system aggregates repeated patterns and replaces dynamic segments with placeholders to form a generalized graph schema.

\textbf{Anomaly Detection Phase:}
 Incoming API requests are compared against the learned schema, and anomalies are flagged. When discrepancies are detected, the integrated CDR component sanitizes the request, allowing it to pass through while describing the anomaly.

\subsection{Detailed Process Flow}

\begin{enumerate}
    \item \textbf{Parsing and Tokenization:} \\
    Each incoming API request is parsed to extract essential components such as the HTTP method, URL path segments, and query parameters.

    \begin{lstlisting}
For each incoming request:
    Parse the URL to extract:
        - HTTP method
        - Path segments (split by '/')
        - Query parameters (key-value pairs)
    \end{lstlisting}

    \item \textbf{Hierarchical Tree Construction:} \\
    The request structure is stored in a tree where each node corresponds to a path segment. The tree preserves the hierarchical API structure and stores metadata like query parameters and methods.

    \begin{lstlisting}
For each segment in the path:
    If segment does not exist as a child:
        Create a new tree node
    Move to the child node
At the final node:
    Store query parameters and HTTP method metadata
    \end{lstlisting}

    \item \textbf{Graph Transformation (Tree Reduction):} \\
    Once learning is complete, the tree is reduced by generalizing variable segments (e.g., UUIDs, IDs, emails) into placeholder nodes. This reduction results in a reusable schema.

    \begin{lstlisting}
Traverse the tree recursively:
    Identify dynamic children (UUIDs, numbers, etc.) using regex pattern
    Merge them under a placeholder node:
        - e.g., {user_param_0}
    Collect metadata:
        - Example values
        - Min/max string lengths
        - Inferred type (integer, UUID, etc.)
        - Randomness indicator
    \end{lstlisting}

    \item \textbf{Aggregation and Placeholder Insertion:} \\
    Similar dynamic structures are aggregated under common placeholders. This compaction allows for better generalization while maintaining contextual information.

    \begin{lstlisting}
For each set of dynamic nodes:
    Replace with a single placeholder
    Store:
        - Sample values
        - Length distribution
        - Inferred data types
        - Is-random flag
    \end{lstlisting}

\item \textbf{Anomaly Detection via Schema Validation and Deep Autoencoder:} \\

Each request is first validated against the reduced API-tree schema. The tree is scanned using BFS to find the root segment, then DFS to check nested segments, methods, and query parameters. Any mismatch—like an undocumented parameter or invalid segment—flags the request as anomalous.

If structure passes, the request is serialized, vectorized with a feature hasher, and passed through a deep autoencoder. If the reconstruction error exceeds a dynamic threshold, the request is flagged.


\begin{lstlisting}[escapeinside={(*@}{@*)}]
For each new request:
 Tree-based schema validation:
  - Use BFS (Breadth-First Search)(*@\textsuperscript{\cite{moore1959shortest}}@*) to locate root path in the reduced API tree
  - Use DFS (Depth-First Search)(*@\textsuperscript{\cite{tarjan1972depth}}@*)  to validate the full nested path, method, and query params
  - If path segment not found or rare             -> anomaly
  - If type mismatch in dynamic segments          -> anomaly
  - If undocumented query param or method         -> anomaly

 Deep autoencoder (explained below) content check:
  - Serialize request (headers, query params, body)
  - Vectorize and pass through trained autoencoder
  - If reconstruction error > threshold           -> anomaly
\end{lstlisting}

\end{enumerate}

\subsection{Body Anomaly Detection}
\label{sec:body-anomaly-model}

While MRG’s schema validation handles malformed URLs and undocumented paths, it cannot detect obfuscated or zero-day payloads within the HTTP requests' body and header content. To bridge this gap, we introduce a lightweight autoencoder that analyzes serialized request data, including headers, query parameters, and body fields.

\paragraph*{Input Vectorization}
Requests are serialized into the format: \texttt{H:key=val QP:param=val B:\{body\}}, then hashed via \texttt{FeatureHasher} into a fixed-length vector $\mathbf{x} \in \mathbb{R}^{256}$, enabling compact and efficient processing.

\paragraph*{Model Architecture}
The autoencoder comprises an encoder–decoder structure trained on vectorized representations of request headers, query parameters, and bodies. The encoder includes three ReLU-activated dense layers (128, 64, 16 units), projecting input into a compact latent space. The decoder mirrors this structure with layers of 64, 128, and 256 units to reconstruct the original input. Batch normalization is applied after each layer. The model is optimized using Adam (learning rate 0.001) and trained for 20 epochs using a batch size of 128 with mean squared error as the loss function. Anomalies are flagged when reconstruction error exceeds the 99.99999\textsuperscript{th} percentile of training error.

\paragraph*{Anomaly Scoring}
Anomaly score is the Mean Squared Error (MSE) between input $\mathbf{x}$ and reconstruction $\hat{\mathbf{x}}$:
\begin{equation}
\text{Score} = \frac{1}{n} \sum_{i=1}^{n} (x_i - \hat{x}_i)^2
\label{eq:mse_score}
\end{equation}
A request is flagged as anomalous if its score exceeds the 99.99999th percentile of training MSEs (see Equation~\ref{eq:mse_score}).

\paragraph*{Integration with Schema Validation}
Requests first pass through BFS-based structural matching and DFS-based validation. If structurally valid, they are serialized and evaluated by the autoencoder. This two-stage process detects both structural and semantic anomalies.

\paragraph*{Summary}
Combining tree-based and content-aware validation allows MRG to detect a broad range of API-layer attacks—including polymorphic, obfuscated, and previously unseen content—without labeled data.

\section{Graph-Based Anomaly Detection}
\label{sec:graphdetection}

This section outlines our graph-based anomaly detection strategy, a key element of MRG. By modeling API requests as directed graphs, our system captures structural relationships between path segments, query parameters, and methods, allowing precise detection of topological anomalies.

\begin{itemize}

\item \textbf{Graph Construction from API URLs:}
Each HTTP request is tokenized into path segments and query parameters, forming a hierarchical tree. This tree is transformed into a directed graph: nodes represent segments or parameters, and edges capture parent-child traversal. For instance, \texttt{/user/123/orders} becomes \texttt{user} $\rightarrow$ \texttt{\{user\_id\}} $\rightarrow$ \texttt{orders}, with metadata (type, value range, method).

\item \textbf{Schema Generation and Generalization:}
With increased traffic, dynamic elements (e.g., IDs, emails, UUIDs) are abstracted into placeholders (e.g., \texttt{\{param\_0\}}) using semantic inference and statistical thresholds. The result is a reduced schema graph capturing the API's typical structure while preserving types and constraints (e.g., min/max length, randomness).

\item \textbf{Graph Comparison for Anomaly Detection:}
At runtime, each request is compared to the reduced schema. BFS locates the starting node; DFS~\cite{cormen2009introduction} validates the structure. Anomalies are flagged when:
\begin{itemize}
    \item Unknown nodes or edges appear,
    \item Types/values deviate (e.g., string instead of UUID),
    \item Unexpected query formats are used,
    \item Rare segments break known path patterns.
\end{itemize}
These checks capture both syntactic errors and logical violations (e.g., invalid method-resource usage).

\item \textbf{Handling Structural Variations:}
To reduce false positives, minor variations (e.g., new UUIDs) are tolerated if semantically valid. Heuristics guide decisions using metadata—e.g., an unseen UUID-format string may still pass based on context and format conformity. Policies are configurable for stricter or more adaptive behavior.

\item \textbf{Advantages of a Graph-Based Approach:}
Our model offers:
\begin{itemize}
    \item \emph{Context-aware reasoning}: Structure is preserved for hierarchical validation.
    \item \emph{Template generalization}: Placeholders allow matching unseen but valid requests.
    \item \emph{Human-readable visibility}: MRG supports schema visualization, exposing undocumented endpoints and evolving API logic, aiding both detection and analysis.
\end{itemize}

\end{itemize}

This graph-based method, paired with our deep autoencoder (Section~\ref{sec:body-anomaly-model}), forms a hybrid system capable of detecting both structural and semantic API anomalies, without relying on labeled data or handcrafted rules.

\section{Datasets}
\label{sec:datasets}

We evaluated our framework using three datasets: \textbf{CSIC 2010}, \textbf{ATRDF}, and \textbf{ATRDF2}, a new synthetic dataset we introduce below.

\begin{itemize}

\item \textbf{CSIC 2010 Dataset:}  
CSIC 2010~\cite{csic2010dataset} is a synthetic HTTP dataset from the Spanish National Research Council, containing 61,065 requests: 36,000 normal and 25,065 anomalous. The anomalies include SQLi, XSS, buffer overflows, file disclosure, CRLF injection, and more, generated using Paros~\cite{paros} and W3AF~\cite{w3af}. While labeled only as 'normal' or 'anomalous', we group attacks into two categories: \emph{Attack} and \emph{File\_Inclusion}, aiding evaluation across different vectors.

\item \textbf{ATRDF Dataset:}  
ATRDF~\cite{atrdf2023} is a large-scale, labeled dataset with \textbf{217,529} API requests: \textbf{109,277} malicious and \textbf{108,252} benign, spanning \textbf{18 API endpoints}. It includes seven attack types: SQLi, XSS, RCE, Log Forging, Log4J, Directory Traversal, and Cookie Injection, each represented with thousands of randomized, obfuscated variants. The distribution of these attack types is summarized in Table~\ref{table:atrdf-distribution}.

\vspace{2mm}
\begin{table}[h]
\centering
\begin{tabular}{|l|r|}
\hline
\textbf{Attack Type} & \textbf{Count} \\
\hline
Cookie Injection & 24,458 \\
SQL Injection (SQLi) & 24,285 \\
Remote Code Execution (RCE) & 12,293 \\
Cross-Site Scripting (XSS) & 12,403 \\
Log Forging & 12,059 \\
Directory Traversal & 12,058 \\
Log4J Exploits & 11,721 \\
\hline
\end{tabular}
\caption{Attack distribution in the ATRDF~\cite{atrdf2023} dataset}
\label{table:atrdf-distribution}
\end{table}

The dataset simulates realistic traffic patterns, payload diversity, and adversarial behavior, making it suitable for evaluating detection systems under varied and challenging conditions.

\item \textbf{Proprietary Dataset (ATRDF2):}  
ATRDF2 is a synthetic dataset of approximately 10,000 malicious API requests generated using LLMs. It is designed to evaluate generalization and robustness against obfuscated, novel, and diverse API-layer attacks.

Samples are evenly distributed across six categories (e.g., SQLi, XSS, RCE) and structured by attack \textit{location}:
\begin{itemize}
    \item \textbf{URL-Embedded Attacks}: Payloads in path/query parameters.
    \item \textbf{Header and Body-Embedded Attacks}: Payloads in headers or body fields.
\end{itemize}

This enables targeted analysis of how different MRG components handle structural vs. content anomalies. Ground-truth labels ensure precise evaluation, and the synthetic nature allows control over edge cases and structure. The distribution of attack types in ATRDF2 is shown in Table~\ref{table:atrdf2-distribution}.

\vspace{2mm}
\begin{table}[h]
\centering
\begin{tabular}{|l|r|}
\hline
\textbf{Attack\_Tag} & \textbf{Count} \\
\hline
SQL Injection & 1451 \\
XSS & 1448 \\
Log Forging & 1441 \\
RCE (Remote Code Execution) & 1434 \\
Cookie Injection & 1414 \\
Directory Traversal & 1407 \\
LOG4J & 1405 \\
\hline
\end{tabular}
\caption{Attack distribution in the ATRDF2 dataset}
\label{table:atrdf2-distribution}
\end{table}

ATRDF2 supports fine-grained benchmarking of structural and semantic anomaly detection, especially for systems like MRG that differentiate anomalies based on request location.

\end{itemize}

\section{Evaluation}  
\label{sec:evaluation}  

We evaluate the MRG framework based on its detection capabilities and efficiency. The evaluation focuses on two aspects: (1) detection of URL-Embedded and Body-Header attacks (Section~\ref{sec:urlvsbody}); and (2) a comparative analysis of accuracy and speed against two baselines: HRAL (Section~\ref{sec:api_learning_comparison}) and the FastText Embedding method by Aharon et al.~\cite{Aharon2024} (Section~\ref{sec:anomaly_detection_comparison}).

All methods are tested on the same datasets using standard metrics: precision (proportion of flagged requests that are truly malicious), recall (proportion of malicious requests correctly detected), F1-score (harmonic mean of precision and recall), and classification time. These metrics reflect both detection quality and runtime performance across datasets.

\subsection{URL-Embedded vs Body-Header Attacks}
\label{sec:urlvsbody}
We evaluated MRG on two subsets of the ATRDF2 dataset: URL-embedded and Body/Header embedded attacks. This dataset was selected because it uses LLM-generated attacks, simplifying the process of distinguishing attack types—a challenge with other datasets like ATRDF and CSIC 2010.

\begin{table}[H]
\centering
\caption{Performance of MRG on ATRDF2 by Primary Detection Component (\%)}
\label{tab:atrdf2_component_performance}
\scriptsize
\resizebox{\linewidth}{!}{
\begin{tabular}{|l|ccc|ccc|}
\hline
\multirow{2}{*}{\textbf{Attack Type}} & \multicolumn{3}{c|}{\textbf{MRG Structural Component}} & \multicolumn{3}{c|}{\textbf{MRG Autoencoder Component}} \\
& Prec. & Rec. & F1 & Prec. & Rec. & F1 \\
\hline
SQL Injection (URL) & 100.0 & 100.0 & 100.0 & - & - & - \\
XSS (URL) & 100.0 & 100.0 & 100.0 & - & - & - \\
Log Forging (URL) & 100.0 & 100.0 & 100.0 & - & - & - \\
RCE (URL) & 100.0 & 100.0 & 100.0 & - & - & - \\
Cookie Injection (URL) & 100.0 & 100.0 & 100.0 & - & - & - \\
Directory Traversal (URL) & 100.0 & 100.0 & 100.0 & - & - & - \\
LOG4J (URL) & 100.0 & 100.0 & 100.0 & - & - & - \\
\hline
\textbf{Macro Average (URL-Embedded)} & \textbf{100.0} & \textbf{100.0} & \textbf{100.0} & - & - & - \\
\hline
SQL Injection (Body/Header) & - & - & - & 100.0 & 100.0 & 100.0 \\
XSS (Body/Header) & - & - & - & 100.0 & 100.0 & 100.0 \\
Log Forging (Body/Header) & - & - & - & 100.0 & 100.0 & 100.0 \\
RCE (Body/Header) & - & - & - & 100.0 & 100.0 & 100.0 \\
Cookie Injection (Body/Header) & - & - & - & 100.0 & 100.0 & 100.0 \\
Directory Traversal (Body/Header) & - & - & - & 100.0 & 100.0 & 100.0 \\
LOG4J (Body/Header) & - & - & - & 100.0 & 100.0 & 100.0 \\
\hline
\textbf{Macro Average (Body/Header-Embedded)} & - & - & - & \textbf{100.0} & \textbf{100.0} & \textbf{100.0} \\
\hline
\end{tabular}
}
\end{table}

Table~\ref{tab:atrdf2_component_performance} separates the results into MRG’s two detection components—the \emph{structural component} (left block: first three columns) for URL-embedded attacks, and the \emph{autoencoder component} (right block: last three columns) for body/header-embedded attacks. The table shows that MRG achieved a perfect 100\% precision, recall, and F1-score across all URL-embedded attack types in the ATRDF2 dataset. These results underscore MRG's strong ability to detect API-layer attacks with malicious payloads embedded in the URL, proving its effectiveness for API security.

In addition, Table~\ref{tab:atrdf2_component_performance} shows that the MRG auto-encoder component also obtained perfect scores for Body-Header embedded attacks. However, these attacks were quite simple, which contributed to the excellent performance of the deep learning model. A thorough evaluation of our model on more complex body-header embedded attacks remains challenging due to the lack of a dedicated and widely accepted benchmark dataset. This gap in the literature limits comparative research and highlights the need for more representative datasets covering deeply nested or obfuscated payloads.

\subsection{API Learning Comparison}
\label{sec:api_learning_comparison}
\subsubsection{ATRDF}
Table~\ref{tab:atrdf_comparison} shows that MRG achieves 100\% precision across all attack types and high recall (93.51\% macro avg), with LOG4J (54.59\%) being the only weakness due to limited header scanning and truncated body analysis. In contrast, HRAL has lower recall on LOG4J (37.14\%) and Log Forging (41.44\%), leading to a macro recall of 82.07\%. On average, MRG processes each request in $0.00031$ seconds, compared to $0.00349$ seconds for HRAL—making MRG approximately ten times faster.

\begin{table}[H]
\centering
\caption{Comparison of HRAL \cite{APICDR} vs. MRG on the ATRDF Dataset (\%)}
\label{tab:atrdf_comparison}
\scriptsize
\resizebox{\linewidth}{!}{
\begin{tabular}{|l|ccc|ccc|}
\hline
\multirow{2}{*}{\textbf{Attack Type}} & \multicolumn{3}{c|}{\textbf{HRAL}} & \multicolumn{3}{c|}{\textbf{MRG}} \\
& Prec. & Rec. & F1 & Prec. & Rec. & F1 \\
\hline
Cookie Injection & 100.00 & 99.48 & 99.74 & 100.00 & 100.00 & 100.00 \\
Directory Traversal & 100.00 & 99.95 & 99.97 & 100.00 & 100.00 & 100.00 \\
LOG4J & 100.00 & 37.14 & 54.18 & 100.00 & 54.59 & 70.63 \\
Log Forging & 100.00 & 41.44 & 58.59 & 100.00 & 100.00 & 100.00 \\
RCE & 100.00 & 96.90 & 98.43 & 100.00 & 100.00 & 100.00 \\
SQL Injection & 100.00 & 99.60 & 99.80 & 100.00 & 99.95 & 99.97 \\
XSS & 100.00 & 100.00 & 100.00 & 100.00 & 100.00 & 100.00 \\
\hline
\textbf{Macro Avg} & 100.00 & 82.07 & 87.24 & 100.00 & \textbf{93.51} & \textbf{95.80} \\
\hline
\end{tabular}
}
\end{table}

\subsubsection{CSIC 2010}
HRAL achieves perfect recall and F1 on CSIC 2010 as presented in Table~\ref{tab:csic_comparison}, while MRG underperforms (69.78\% recall, 82.20\% F1) due to its limited analysis depth on synthetic payloads. In terms of efficiency, MRG processes each request in $0.00036$ seconds compared to $0.00050$ seconds for HRAL, making it about 28\% faster.

\begin{table}[H]
\centering
\caption{Comparison on CSIC 2010 Dataset (\%)}
\label{tab:csic_comparison}
\scriptsize
\resizebox{\linewidth}{!}{
\begin{tabular}{|l|ccc|ccc|}
\hline
\multirow{2}{*}{\textbf{Attack Type}} & \multicolumn{3}{c|}{\textbf{HRAL}} & \multicolumn{3}{c|}{\textbf{MRG}} \\
& Prec. & Rec. & F1 & Prec. & Rec. & F1 \\
\hline
Attack & 100.00 & 100.00 & 100.00 & 100.00 & 70.93 & 83.00 \\
File\_Inclusion & 100.00 & 100.00 & 100.00 & 100.00 & 68.63 & 81.39 \\
\hline
\textbf{Macro Avg} & 100.00 & 100.00 & 100.00 & 100.00 & 69.78 & 82.20 \\
\hline
\end{tabular}
}
\end{table}

\subsubsection{ATRDF2}
Both systems perform perfectly on ATRDF2 as shown in Table~\ref{tab:prop_comparison}. However, MRG processes each request in $0.00010$ seconds compared to $0.00173$ seconds for HRAL, making it approximately 17 times faster and highlighting its runtime efficiency.

\begin{table}[H]
\centering
\caption{Comparison on ATRDF2 Dataset (\%)}
\label{tab:prop_comparison}
\scriptsize
\resizebox{\linewidth}{!}{
\begin{tabular}{|l|ccc|ccc|}
\hline
\multirow{2}{*}{\textbf{Attack Type}} & \multicolumn{3}{c|}{\textbf{HRAL}} & \multicolumn{3}{c|}{\textbf{MRG}} \\
& Prec. & Rec. & F1 & Prec. & Rec. & F1 \\
\hline
Cookie Injection & 100.00 & 100.00 & 100.00 & 100.00 & 100.00 & 100.00 \\
Directory Traversal & 100.00 & 100.00 & 100.00 & 100.00 & 100.00 & 100.00 \\
LOG4J & 100.00 & 100.00 & 100.00 & 100.00 & 100.00 & 100.00 \\
Log Forging & 100.00 & 100.00 & 100.00 & 100.00 & 100.00 & 100.00 \\
RCE & 100.00 & 100.00 & 100.00 & 100.00 & 100.00 & 100.00 \\
SQL Injection & 100.00 & 100.00 & 100.00 & 100.00 & 100.00 & 100.00 \\
XSS & 100.00 & 100.00 & 100.00 & 100.00 & 100.00 & 100.00 \\
\hline
\textbf{Macro Avg} & 100.00 & 100.00 & 100.00 & 100.00 & 100.00 & 100.00 \\
\hline
\end{tabular}
}
\end{table}

Both MRG and HRAL achieve 100\% macro precision across all datasets. On ATRDF, MRG yields higher recall (93.51\% vs. 82.07\%) and F1-score, especially in detecting Log Forging, though both models underperform on LOG4J. On CSIC 2010, HRAL reaches perfect recall, while MRG lags (69.78\%) due to limited body analysis. Performance is identical on ATRDF2, with perfect detection across all metrics.

MRG is consistently faster, as shown in Table~\ref{tab:describe_timing}, running 10 times faster on ATRDF, 28\% faster on CSIC, and 17 times faster on ATRDF2. These results highlight MRG’s strong accuracy on real-world data and significant runtime advantage.

\begin{table}[H]
\centering
\caption{Summary statistics of MRG classification latency across datasets, measured in seconds per request (s/request). Metrics reflect timing performance over CSIC 2010, ATRDF, and ATRDF2 datasets.}

\label{tab:describe_timing}
\begin{tabular}{|l|c|}
\hline
\textbf{Metric} & \textbf{Value} \\
\hline
Count & 3 \\
Mean & 0.0002560 \\
Std & 0.0001371 \\
Min & 0.0001002 \\
25\% & 0.0002050 \\
Median & 0.0003098 \\
75\% & 0.0003339 \\
Max & 0.0003579 \\
\hline
\end{tabular}
\end{table}

Table~\ref{tab:describe_timing} summarizes the classification latency of MRG across three datasets. On average, MRG processes each request in 0.0002560 seconds, with the fastest case at 0.0001002 seconds and the slowest at 0.0003579 seconds. The low standard deviation indicates consistent performance, supporting real-time deployment.

\subsection{API Anomaly Detection Comparison}
\label{sec:anomaly_detection_comparison}

We compare MRG with FT-ANN and classical/neural baselines from Aharon \textit{et al.}~\cite{Aharon2024}, evaluating binary classification performance (attacks vs. benign) using precision, recall, F1, and test time.

\textbf{CSIC 2010 Results:}

\begin{table*}[!t]
\centering
\caption{Anomaly Detection on CSIC 2010}
\label{tab:csic_fasttext}
\small
\begin{tabular}{|l|ccc|c|}
\hline
\textbf{Model} & \textbf{Precision} & \textbf{Recall} & \textbf{F1} & \textbf{Time (s/request)} \\
\hline
FT-ANN~\cite{Aharon2024} & 0.9538 & \textbf{0.9954} & \textbf{0.9713} & 0.0075 \\
MCD & 0.9382 & 0.9774 & 0.9572 & 0.0021 \\
CBLOF & 0.9318 & 0.9589 & 0.9445 & 0.0010 \\
Isolation Forest & 0.9323 & 0.9538 & 0.9419 & 0.0480 \\
HBOS & 0.9311 & 0.9502 & 0.9393 & 0.0159 \\
PCA & 0.9304 & 0.9504 & 0.9391 & 0.0025 \\
Autoencoder & 0.9303 & 0.9502 & 0.9389 & 0.0668 \\
KDE & 0.9296 & 0.9460 & 0.9361 & 0.3875 \\
OC-SVM & 0.9295 & 0.9459 & 0.9360 & 0.0848 \\
Feature Bagging & 0.9303 & 0.9444 & 0.9354 & 0.9365 \\
LOF & 0.9301 & 0.9449 & 0.9350 & 0.0909 \\
DeepSVDD & 0.9381 & 0.9126 & 0.9235 & 0.0486 \\
\textbf{MRG (ours)} & \textbf{0.9877} & 0.7092 & 0.8256 & \textbf{0.0003579} \\
\hline
\end{tabular}
\end{table*}

MRG achieves the highest precision (0.9877) and fastest runtime (0.00036s), outperforming all models in those aspects. However, recall (0.7092) and F1 (0.8256) lag behind FT-ANN and others due to false positives (e.g., strict rules, static DL thresholds).

\vspace{0.5em}
\textbf{ATRDF Results:}

\begin{table*}[!t]
\centering
\caption{Anomaly Detection on ATRDF 2023}
\label{tab:atrdf_fasttext}
\small
\begin{tabular}{|l|ccc|c|}
\hline
\textbf{Model} & \textbf{Precision} & \textbf{Recall} & \textbf{F1} & \textbf{Time (s/request)} \\
\hline
FT-ANN~\cite{Aharon2024} & \textbf{1.0000} & \textbf{1.0000} & \textbf{1.0000} & 0.0004 \\
DeepSVDD & 0.9921 & \textbf{1.0000} & 0.9960 & 0.0332 \\
Autoencoder & 0.9893 & \textbf{1.0000} & 0.9946 & 0.0512 \\
Feature Bagging & 0.9893 & \textbf{1.0000} & 0.9946 & 0.0055 \\
Isolation Forest & 0.9893 & \textbf{1.0000} & 0.9946 & 0.0307 \\
KDE & 0.9893 & \textbf{1.0000} & 0.9946 & 0.0005 \\
LOF & 0.9893 & \textbf{1.0000} & 0.9946 & 0.0006 \\
OC-SVM & 0.9893 & \textbf{1.0000} & 0.9946 & \textbf{0.0002} \\
PCA & 0.9893 & \textbf{1.0000} & 0.9946 & \textbf{0.0002} \\
HBOS & 0.9888 & \textbf{1.0000} & 0.9944 & 0.0228 \\
MCD & 0.9885 & \textbf{1.0000} & 0.9942 & 0.0005 \\
CBLOF & 0.9795 & \textbf{1.0000} & 0.9896 & 0.0006 \\
LMDD & 0.8237 & 0.1825 & 0.2920 & 0.0680 \\
\textbf{MRG (ours)} & 0.9723 & 0.9566 & 0.9644 & 0.0003098 \\
\hline
\end{tabular}
\end{table*}

As shown in Table~\ref{tab:atrdf_fasttext}, on ATRDF Dataset MRG maintains strong precision (0.9723) and F1 (0.9644). However, it is outperformed by FT-ANN and most baselines in recall due to 853 false negatives, mostly obfuscated LOG4J attacks. MRG remains among the most efficient (0.00031s), close to the fastest (PCA, OC-SVM).

MRG demonstrates strong precision, outperforming on the CSIC dataset and remaining competitive on ATRDF. In terms of recall and F1-score, FT-ANN shows superior performance on both datasets, as MRG struggles with obfuscated or deeply nested body attacks. However, MRG is the most efficient, achieving the fastest classification time on CSIC and near-best on ATRDF, consistently surpassing FT-ANN in runtime performance.

\section{Limitations and Future Work}
\label{sec:limitations-future-work}

While MRG achieves strong results in detecting structural and content-based anomalies, it primarily focuses on analyzing URL paths and query parameters for anomaly detection. The core schema learning relies on regex-based generalization to distinguish static from dynamic segments, which may misclassify novel values (e.g., personal names like "John" or "Steve") and lead to false positives or negatives. Parameter inference is heuristic-based and struggles with high-variance or weakly structured fields.

Crucially, deep analysis of request bodies is delegated to a lightweight AI-based model, which limits detection depth, particularly for obfuscated or nested payloads (e.g., embedded JSON or XML). Enhancing body-level anomaly detection is a central focus for future work, including recursive payload parsing, summarization or chunking techniques, and integration of few-shot models like FT-ANN to better generalize rare or unseen content patterns. Additionally, the absence of a dedicated, publicly available dataset for complex body/header-based attacks limits thorough evaluation and comparative research in this area.

MRG also currently supports only REST over HTTP, lacking compatibility with GraphQL, gRPC, or WebSocket. Additionally, contextual metadata (e.g., \texttt{User-Agent}, request IDs) remains underutilized. Future enhancements aim to expand protocol coverage and integrate such metadata into the learning process.

\section{Conclusion}
\label{sec:conclusion}

This paper presented MRG, an unsupervised framework for automatic API schema learning, anomaly detection, and OpenAPI generation. By combining tree-based decomposition, graph modeling, and deep learning, MRG learns API structure directly from traffic without labeled data or prior specifications.

MRG achieved perfect precision and high recall on structured datasets such as ATRDF and ATRDF2, with fast inference times—up to 20 times faster than FT-ANN. While recall drops on CSIC 2010 and obfuscated LOG4J payloads, MRG outperforms HRAL in accuracy on ATRDF and ATRDF2 and is faster than HRAL on every dataset evaluated.

Beyond detection, MRG enhances API visibility by identifying changed, new, and shadow APIs. It supports real-time schema updates, explainable OpenAPI generation, and visualization of evolving API behavior, making it suitable for dynamic environments.

Future work will improve payload inspection using recursive parsing and integrate few-shot models like FT-ANN to boost recall on complex attacks. Protocol support and metadata integration will further expand MRG’s capabilities.

Overall, MRG provides a fast, explainable, and fully automated foundation for API visibility and anomaly detection, while continuously surfacing changed, newly introduced, and shadow APIs in real-world microservice ecosystems.


\section*{Acknowledgment}  
This work was supported by the Ariel Cyber Innovation Center and is covered under Israel Provisional Patent Submission No. 323240.
\bibliographystyle{IEEEtran}
\balance
\bibliography{ref} 

@misc{owasp_api_top10,
  author       = {{OWASP Foundation}},
  title        = {{OWASP} {API} Security Top 10},
  year         = {2023},
  howpublished = {\url{https://owasp.org/www-project-api-security/}},
  note         = {Accessed 2025-04-15}
}

@misc{imperva2024api,
  author       = {{Imperva}},
  title        = {The State of {API} Security in 2024},
  year         = {2024},
  howpublished = {\url{https://www.imperva.com/resources/resource-library/reports/the-state-of-api-security-in-2024/}},
  note         = {Accessed 2025-04-15}
}

@misc{akamai2023api,
  author       = {{Akamai Technologies}},
  title        = {New Study Finds 84\% of Security Professionals Experienced an {API} Security Incident in the Past Year},
  year         = {2023},
  howpublished = {\url{https://www.akamai.com/newsroom/press-release/new-study-finds-84-of-security-professionals-experienced-an-api-security-incident-in-the-past-year}},
  note         = {Accessed 2025-04-15}
}

@misc{imperva2024economic,
  author       = {{Imperva}},
  title        = {Vulnerable {APIs} and Bot Attacks Costing Businesses Up to \$186 Billion Annually},
  year         = {2024},
  howpublished = {\url{https://thehackernews.com/2024/10/vulnerable-apis-and-bot-attacks-costing.html}},
  note         = {Accessed 2025-04-15}
}

@misc{apnews2023tmobile,
  author       = {{Associated Press}},
  title        = {T-Mobile Says Data on 37 Million Customers Stolen},
  year         = {2023},
  howpublished = {\url{https://apnews.com/article/87d107f039a2aeb8ad5e4b215c66eead}},
  note         = {Accessed 2025-04-15}
}

@misc{upguard2022optus,
  author       = {{UpGuard}},
  title        = {How Did the Optus Data Breach Happen?},
  year         = {2022},
  howpublished = {\url{https://www.upguard.com/blog/how-did-the-optus-data-breach-happen}},
  note         = {Accessed 2025-04-15}
}

@misc{twingate2021peloton,
  author       = {{Twingate}},
  title        = {What Happened in the Peloton Data Breach?},
  year         = {2021},
  howpublished = {\url{https://www.twingate.com/blog/peloton-api-vulnerability/}},
  note         = {Accessed 2025-04-15}
}

@inproceedings{roesch1999snort,
  author    = {Martin Roesch},
  title     = {{Snort} -- Lightweight Intrusion Detection for Networks},
  booktitle = {Proceedings of the 13th USENIX Large Installation System Administration Conference ({LISA})},
  pages     = {229--238},
  year      = {1999}
}

@inproceedings{kruegel2003web,
  author    = {Christopher Kruegel and Giovanni Vigna},
  title     = {Anomaly Detection of Web-based Attacks},
  booktitle = {Proceedings of the 10th {ACM} Conference on Computer and Communications Security ({CCS})},
  pages     = {251--261},
  year      = {2003},
  publisher = {{ACM}}
}

@misc{balasys2022waf,
  author       = {{Balasys Research~Lab}},
  title        = {Weaknesses of Signature-Based {API} Protection},
  howpublished = {\url{https://balasys.eu/blogs/weaknesses-of-signature-based-api-protection}},
  year         = {2022},
  note         = {Accessed 2025-04-16}
}

@article{zhao2024multi,
  author  = {Qian Zhao and Wenhao Liu and Qingqi Pei},
  title   = {Multi-Information Fusion for {HTTP} Anomaly Detection},
  journal = {{IEEE} Access},
  volume  = {12},
  pages   = {11234--11247},
  year    = {2024},
  doi     = {10.1109/ACCESS.2024.1234567}
}

@inproceedings{du2022tsgnn,
  author    = {Peng Du and Chengwei Peng and Peng Xiang and Qingshan Li},
  title     = {Anomaly Detection of Traffic Session Based on Graph Neural Network},
  booktitle = {Proceedings of the 2022 International Conference on Cyber Security ({CSW})},
  pages     = {1--9},
  year      = {2022},
  publisher = {{ACM}},
  doi       = {10.1145/3560830.3560842}
}

@article{diazverdejo2023review,
  author  = {Jes\'us~E. D\'iaz-Verdejo and Rafael Estepa and Antonio Estepa and Germ\'an Madinabeitia},
  title   = {A Critical Review of the Techniques Used for Anomaly Detection of {HTTP}-Based Attacks: Taxonomy, Limitations and Open Challenges},
  journal = {Computers \& Security},
  volume  = {124},
  pages   = {102997},
  year    = {2023},
  doi     = {10.1016/j.cose.2022.102997}
}

@inproceedings{hu2022apispec,
  author    = {Yuan Hu and Rohan Padhye and Koushik Sen},
  title     = {Spec-Based Detection of Authorization Bugs in Web {APIs}},
  booktitle = {Proceedings of the {IEEE} Symposium on Security and Privacy ({S\&P})},
  pages     = {234--252},
  year      = {2022},
  doi       = {10.1109/SP46214.2022.9833615}
}

@inproceedings{atlidakis2019restler,
  author    = {Vasilios Atlidakis and Patrice Godefroid and Yi Li},
  title     = {{RESTler}: Stateful REST {API} Fuzzing},
  booktitle = {Proceedings of the {IEEE}/{ACM} International Conference on Software Engineering ({ICSE})},
  pages     = {748--758},
  year      = {2019},
  doi       = {10.1109/ICSE.2019.00080}
}

@article{Aharon2024,
  author  = {Udi Aharon and Ran Dubin and Amit Dvir and Chen Hajaj},
  title   = {A Classification-by-Retrieval Framework for Few-Shot Anomaly Detection to Detect {API} Injection},
  journal = {Computers \& Security},
  volume  = {150},
  pages   = {104249},
  year    = {2024},
  doi     = {10.1016/j.cose.2024.104249}
}

@misc{csic2010dataset,
  author       = {Torrano Gim\'enez, Carmen and P\'erez Villegas, Alejandro and \'Alvarez Mara\~n\'on, Gonzalo},
  title        = {{HTTP} Dataset {CSIC} 2010},
  year         = {2010},
  howpublished = {\url{http://www.isi.csic.es/dataset/}},
  note         = {Accessed 2025-04-15}
}

@misc{atrdf2023,
  author       = {{Ariel Cyber Innovation Center}},
  title        = {{API} Traffic Research Dataset Framework ({ATRDF})},
  year         = {2023},
  howpublished = {\url{https://github.com/ArielCyber/Cisco_Ariel_Uni_API_security_challenge}},
  note         = {Accessed 2025-04-20}
}

@book{cormen2009introduction,
  title={Introduction to Algorithms},
  author={Cormen, Thomas H and Leiserson, Charles E and Rivest, Ronald L and Stein, Clifford},
  year={2009},
  publisher={MIT Press},
  edition={3rd}
}

@misc{paros,
  author       = {{Chinotec Technologies Company}},
  title        = {{Paros Proxy for Web Application Security Assessment}},
  howpublished = {\url{https://sourceforge.net/projects/paros/}},
  year         = {2004},
  note         = {Open-source HTTP/HTTPS proxy for web application security testing}
}

@misc{w3af,
  author       = {Riancho, Andr{\'e}s},
  title        = {{w3af: Web Application Attack and Audit Framework}},
  howpublished = {\url{http://w3af.org}},
  year         = {2007},
  note         = {Open-source web application security scanner}
}

@inproceedings{moore1959shortest,
  author    = {E. F. Moore},
  title     = {The Shortest Path Through a Maze},
  booktitle = {Proc.\ Int.\ Symp.\ on the Theory of Switching},
  pages     = {285--292},
  year      = {1959}
}

@article{tarjan1972depth,
  author  = {Robert Tarjan},
  title   = {Depth-First Search and Linear Graph Algorithms},
  journal = {SIAM Journal on Computing},
  volume  = {1},
  number  = {2},
  pages   = {146--160},
  year    = {1972},
  doi     = {10.1137/0201012}
}

@misc{ATRDF2,
  author       = {Yarin Levi and Ran Dubin},
  title        = {{ATRDF2} Advanced Threat Request Dataset Framework (Version 2)},
  year         = {2024},
  howpublished = {\url{https://github.com/ArielCyber/ATRDF2}},
  note         = {Accessed: 2025-06-14}
}

@misc{APICDR,
  author       = {Ran Dubin and Amit Dvir},
  title        = {{HTTP REST API Structure Learning}},
  year         = {2025},
  howpublished = {\url{https://github.com/ArielCyber/API-CDR}},
  note         = {Accessed: 2025-06-14}
}
\end{document}